# MODELING OF USER PORTRAIT THROUGH SOCIAL MEDIA

*Haiqian Gu[1,2], Jie Wang[3], Ziwen Wang[1,2], Bojin Zhuang[3], Fei Su[1,2]*

[1]School of Information and Communication Engineering
[2]Beijing Key Laboratory of Network System and Network Culture
Beijing University of Posts and Telecommunications, Beijing, China
[3]Ping An Technology (Shenzhen) Co., Ltd.
mixiu@bupt.edu.cn, wangjie388@pingan.com.cn, wangziwen@bupt.edu.cn,
zhuangbojin232@pingan.com.cn, sufei@bupt.edu.cn

**ABSTRACT**

Nowadays, massive useful data of user information and social behavior have been accumulated on the Internet, providing a possibility of profiling user's personality traits online. In this paper, we propose a psychological modeling method based on computational linguistic features to profile Big Five personality traits of users on Sina Weibo (a Twitter-like microblogging service in China) and their correlations with user's social behaviors. To the best of our knowledge, this is the first research on investigating the potential relationship between profile information, social-network behaviors and personality traits of users on Sina Weibo. Our results demonstrate an effective modeling approach to understanding demographic and psychological portraits of users on social media without customer disruption, which is useful for commercial incorporations to provide better personalized products and services.

## 1. INTRODUCTION

Personality has been widely studied as it both reflects and affects people's behavior, which exhibits promising applications in precision marketing. Recently, data-driven psychological interpretation of users' personality has drawn much attention. Sina Weibo, as a very popular and important social media in China, are posted about 100 million microblogs every day [1]. With accumulation of such large-scale online data, conventional questionnaire-based personality measurement becomes expensive and inefficient.

To address this issue, various data-driven modeling methods have been studied for user portrait computation based on user information and social behavior. Zhao et al. [2] proved the validity of SC-LIWC (Simplified Chinese version of Linguistic Inquiry and Word Count) in detecting psychological expressions in SNS short texts and Qiu et al.[3] studied the relationship between Chinese short texts and corresponding word categories of SC-LIWC. These two results are indeed fundamental research foundation for our work. Furthermore, Golbeck et al. [4] demonstrated that public information shared on Facebook could be used to predict users' Big Five personality. Sumner et al. [5] demonstrated that there were some links between Dark Triad constructs and Twitter usage and employed a variety of machine learning techniques to predict these constructs in users. However, both [4] and [5] ignored users' social behaviors, and our study make up for this limitation. On the other side, Hung et al. [6] introduced a tag-based user profiling for social media recommendation, but they did not give concrete user profiles. In short, most of the previous researches on this area utilized only one type of information on social network, such as text data or profile features. In our study, we take a comprehensive consideration on correlations between Big Five personality trait dimensions and user's information and social behavior, and successfully predict user's personality.

The key contributions of this paper are summarized as follows:
- An effective calculation model of Big Five personality scores based on SC-LIWC word frequencies has been proposed and demonstrated, personality profiles of Sina Weibo users have been calculated;
- 6,467 valid Weibo users have been selected to compute Big Five personality scores based on their posted short texts. And the way we train the personality calculation model has also been discussed;
- Correlation analysis between Big Five personality scores and SC-LIWC word frequency features, user's tags, user's demographics, user's emoticon usage and user's behaviors has comprehensively conducted, verifying the possibility of profiling user's personality from microblog texts and other information they share on Sina Weibo.

Thanks to Chinese National Natural Science Foundation (61532018, 61471049) for funding.

978-1-5386-1737-3/18/$31.00 ©2018 IEEE

## 2. THEORETICAL BACKGROUND

### 2.1. The Big Five personality

The Big Five model consisting of five categorical personality traits, i.e. Openness, Conscientiousness, Extraversion, Agreeableness and Neuroticism, has become one of the most widely-adopted psychological analysis models recently [7].

The Big Five model is characterized as shown in Table 1 [8]. It has well examined and developed as an important psychometric method by many researchers, which providing a comprehensive profile of an individual's cognitive patterns.

**Table 1.** The dimensions of Big Five model

| Dimension | Score | Personal traits |
|---|---|---|
| Openness | high | Wide interests, Imaginative, Intelligent, Curious |
|  | low | Commonplace, Simple, Shallow, Unintelligent |
| Conscientiousness | high | Organized, Tend to plan, Efficient, Responsible |
|  | low | Careless, Disorderly, Frivolous, Irresponsible |
| Extroversion | high | Talkative, Active, Energetic, Enthusiastic |
|  | low | Quiet, Reserved, Shy, Silent |
| Agreeableness | high | Sympathetic, Kind, Appreciative, Generous |
|  | low | Fault-finding, Cold, Unfriendly, Cruel |
| Neuroticism | high | Tense, Anxious, Nervous, Worried |
|  | low | Stable, Calm, Contented, Unemotional |

### 2.2. SC-LIWC

Linguistic Inquiry and Word Count (LIWC) [9] is a valid text analysis model based on word counts of psychologically meaningful categories. However, it was first developed only in English. To meet the demand of processing Simplified Chinese texts, Gao et al. [10] developed a Simplified Chinese version of LIWC (SC-LIWC) based on early version of LIWC and its traditional Chinese version (C-LIWC) [11]. Moreover, high frequency words in Chinese social networks has been added into the lexicon of SC-LIWC for better analysis of Sina Weibo short texts. For this study, SC-LIWC is used to calculate word frequency features .

## 3. METHODOLOGY

### 3.1. Personality data collection

For data preparation, we have collected effective Big Five personality scores of 100 Weibo volunteers online. Furthermore, most recent 200 microblogs of 9,555 Weibo users have been collected, including those 100 volunteers.

Additionally, other profile information of users, such as age, gender, location, education and so on have been collected simultaneously. Based on specific filtering criteria, invalid users including marketing and inactive accounts have been excluded. For example, certain accounts whose posts contain advertisement URLs and number of followers is less than 10. Reposted microblogs have been collected in the same way as original ones based on the hypothesis of users' agreements on those reposted texts. In total, social data of 6,467 valid users has been used in this paper.

### 3.2. Data cleaning

In order to obtain high-quality microblog text, a variety of preprocessing and basic natural language processing work has been done beforehand, such as word segmentation and text cleaning. The implementation details of preprocessing are described as follows:

1. Remove URL links (e.g. http://…), Weibo user names (e.g. @username-with symbol @ indicating a user name), hashtags (e.g. #example#), Weibo special words (e.g. reply, repost) and geo-locations;
2. Remove text generated by the system automatically (e.g. Sorry this microblog had been deleted);
3. Remove the advertisements and spam message including certain key words such as "Taobao" (a consumer-to-consumer retail platform in China);
4. Segment Chinese Word with "Jieba" (a Chinese text segmentation tool) to generate a sequence of words.

### 3.3. Mapping matrix

With the help of Scikit-learn module in Python, we calculate the mapping matrix between Big Five personality scores and SC-LIWC word frequencies through the true Big Five personality scores of the 100 volunteers and SC-LIWC word frequencies calculated by their microblog texts. Thus the psychological calculation model is established.

## 4. RAW DATA STATISTICS

### 4.1. Summary of demographic information

Inevitably, people may share fake information or just keep the default choice. For example, we find many people were born on "1970-01-01" which is Weibo's default birthday option. To eliminate the influence of fake information as much as possible, we limit the range of age from 10 to 47(if a user keeps default birthday option, then his age will be 48).

To have an overall view of the data, we draw Table 2 to show various demographic statistics from 6,467 valid users, presenting a diverse population of participants.

**Table 2.** Demographic statistics of 6,467 users

| Items | Percentage | Values |
|---|---|---|
| Age | | 10-47 years (Mean = 26 years, S.D. = 5 years) |
| Gender | 54.3% | Female |
| | 45.7% | Male |
| Verified | 25.7% | Verified |
| | 74.3% | Unverified |
| Tags | 99.2% | Shared (Mean = 6 labels, S.D. = 3 labels) |
| | 0.8% | Unknown |
| Location | 92.8% | Shared |
| | 7.2% | Unknown |
| University | 55.9% | Shared |
| | 44.1% | Unknown |

**Table 3.** 6,467 users' average Big Five scores and standard deviation (S.D.)

| | Mean | S.D. |
|---|---|---|
| Openness | 47.8 | 16.8 |
| Conscientiousness | 49.4 | 13.3 |
| Extroversion | 45.4 | 13.8 |
| Agreeableness | 55.1 | 10.1 |
| Neuroticism | 50.4 | 12.1 |

## 4.2. Summary of Big Five personality scores

Based on the mapping matrix, we calculate 6,467 users' Big Five personality scores and draw Table 3 to describe average values for each personality trait dimension among all users.

## 4.3. Distribution of user tags

Weibo users can optionally write up to 10 tags to show their interests or job fields. Among the 6,467 valid users, there are 5,656 users who tag themselves with several labels, and we have collected 9,620 different tags in total. However, tags which have been quoted by more than 10 users account for only 4%. This is because the tags are written in user's own words, contributing to the large diversity. Anyway, the most frequent tags are still very typical.

From a tag-based user profile, we may know a user is a "post-90s", addicted to "Music" and "Traveling", and is a "humorous" boy.

## 5. ANALYSIS RESULTS

### 5.1. Correlations between personality scores and SC-LIWC word frequencies

We perform *Pearson* correlation values between SC-LIWC word frequencies and the Big Five personality scores. Table 4 lists the SC-LIWC word categories significantly correlated with at least one personality trait. Strong correlations are highlighted in bold. All values shown in Table 4 are reliable for $p < 0.05$, except for some unreliable values with trailing asterisks. Many correlations between users' word use and Big Five personality scores have been found.

**Table 4.** Pearson correlation values between SC-LIWC word frequencies and Big Five personality scores.

| LIWC | O | C | E | A | N |
|---|---|---|---|---|---|
| I (本人、自己、我) | 0.136 | -0.099 | **0.244** | -0.040 | 0.056 |
| We (我们、我俩、咱们) | -0.034 | 0.044 | 0.069 | **0.205** | 0.033 |
| They (他们、她们) | -0.072 | -0.130 | -0.085 | **0.227** | 0.063 |
| Verb (分享、做、提出) | 0.177 | -0.002* | **0.264** | 0.194 | 0.075 |
| Quant (一些、众多、所有) | 0.013* | **0.239** | 0.060 | 0.142 | 0.051 |
| SpecArt (本、该、每) | -0.020* | **0.166** | -0.001* | 0.048 | **-0.216** |
| Social (给、打电话、见面) | 0.157 | -0.054 | **0.206** | **0.234** | 0.086 |
| Affect (怜悯、温暖、敏锐) | 0.164 | 0.076 | **0.220** | 0.088 | -0.114 |
| PosEmo (信心、满足、祝福) | 0.180 | 0.126 | **0.236** | 0.030 | -0.140 |
| NegEmo (担忧、猜疑、报复) | 0.049 | **-0.103** | -0.005* | -0.009* | -0.044 |
| Anx (不安、顾虑、怀疑) | **-0.244** | **-0.290** | -0.088 | -0.132 | 0.047 |
| Ingest (吃、喝、口渴) | -0.035 | -0.133 | -0.043 | 0.031 | **0.302** |
| Achieve (擅长、挑战、胜利) | -0.098 | 0.196 | -0.089 | **0.250** | -0.093 |
| Love (爱、接吻、表白) | 0.124 | 0.095 | **0.250** | 0.091 | -0.004* |
| Hear (说话、声音、呐喊) | -0.027 | -0.052 | 0.190 | 0.003* | **-0.238** |

The Extraversion is associated with words "I" (r =0.244), while the Agreeableness is always related to words "We" (r =0.205).

Individuals using a lot of "Verbs" tend to be extraverted (r =0.264). Those who prefer to use many "Social" words, not surprisingly, have a distinct positive relationship with Extroversion (r =0.206) and Agreeableness (r =0.234).

The frequency of Quant words is strongly correlated with Conscientiousness (r =0.302), which suggests that conscientious people prefer to describing things with measure words. Some special articles in Chinese, such as "should", are positively related with Conscientiousness (r =0.166), while negatively related with Neuroticism (r =-0.216).

"Affect" and "Love" words describing feelings, and positive emotion words ("PosEmo") are positively correlated with Agreeableness (r =0.220, 0.250 and 0.236 respectively). However, the frequency of words that express anxiety ("Anx") negatively correlates with Openness and Conscientiousness (r =-0.244 and -0.290 respectively).

Another interesting find is that the "Ingest" words are strongly correlated with Neuroticism (r =0.264).

Based on our study, we come to the conclusion that word use is indeed an important reflection of individual personality.

### 5.2. Correlations between personality and tags

On each dimension of the Big Five model, we separate users into two polarity groups: high score group (the highest 25%) and low score group (the lowest 25%) to study the differences in their tags. Figure 1 (a) - (j) vividly show the top related tags for opposite personality traits.

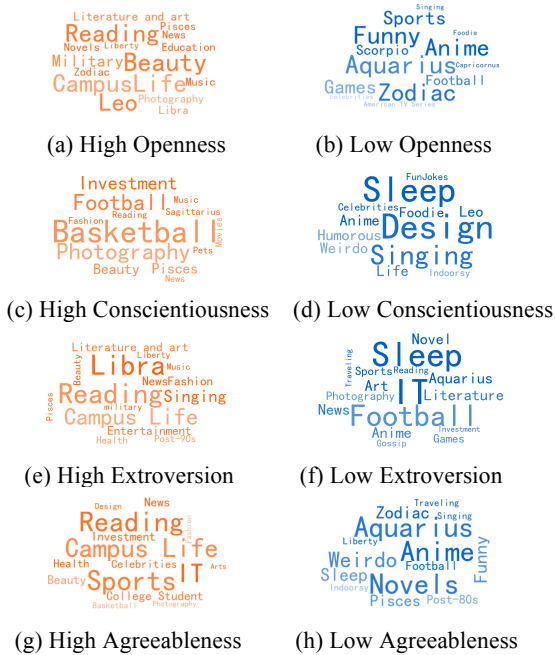

(a) High Openness (b) Low Openness
(c) High Conscientiousness (d) Low Conscientiousness
(e) High Extroversion (f) Low Extroversion
(g) High Agreeableness (h) Low Agreeableness

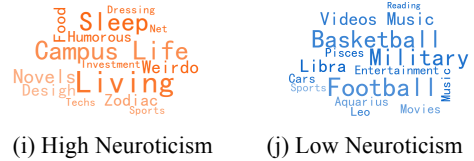

(i) High Neuroticism (j) Low Neuroticism

**Fig. 1.** Word clouds of top related tags on opposite Big Five personality

Comprehensively analyzing the five dimensions, we can draw simple user portrait for certain personality trait group or certain tag. For instance, users tagged with "Sleep" usually score high in Neuroticism, while score low in Agreeableness, and Extroversion.

### 5.3. Correlations between user's demographic information and personality

We analyze the correlation between user's age and Big Five personality scores. As is shown in Figure 2, users obviously score higher in Conscientiousness and Agreeableness with increasing age. Learning how to deal with problems and to be more friendly is part of maturing. On the contrary, users seem to score lower in Openness and Extroversion with increasing age. Age limits people's willing of accepting new things. There are not many changes in Neuroticism, which suggests Neuroticism may be something stable in personality trait.

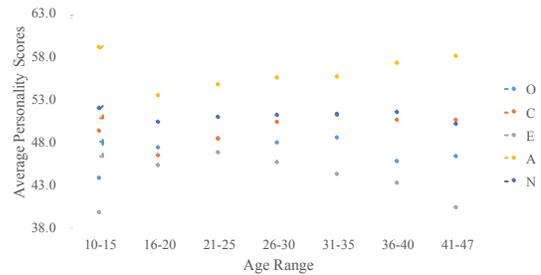

**Fig. 2.** The correlations between age and personality scores

Gender difference is mainly embodied in Extroversion, Conscientiousness and Neuroticism. Men are generally more conscientious, while women are more extraverted and neurotic (shown in Table 5).

**Table 5.** The average Big Five scores between men and women

|  | O | C | E | A | N |
|---|---|---|---|---|---|
| Men | 48.0 | **50.1** | 44.4 | 55.4 | 49.7 |
| Women | 47.6 | 48.7 | **46.5** | 54.9 | **51.6** |

## 5.4. Correlations between user's behavior and personality

### 5.4.1. Verified information and personality

The Sina Weibo system offers a way to personal verified. According to our study, the verified users score higher in Conscientiousness and Neuroticism, while the users who haven't verified score higher in Extroversion (shown in Table 6). Verified users tend to have more fans and have more influence to the society, which may account for why they are more conscientious and neurotic.

**Table 6.** The average Big Five scores between verified user group and unverified user group

|  | O | C | E | A | N |
|---|---|---|---|---|---|
| Verified | 47.2 | **49.9** | 43.8 | 54.8 | **51.9** |
| Unverified | 47.9 | 48.3 | **45.7** | 55.2 | 50.4 |

### 5.4.2. Educational information and personality

Users can optionally fill in their educational information from primary school to university, even their job information.

According to our study, the group who share their educational information score higher in Conscientiousness and Agreeableness (shown in Table 7).

**Table 7.** The average Big Five scores between shared educational information group and unknown group

|  | O | C | E | A | N |
|---|---|---|---|---|---|
| Shared | 47.8 | **49.7** | 45.5 | **55.4** | 50.7 |
| Unknown | 47.8 | 49.0 | 45.2 | 54.8 | 50.6 |

Another interesting find is that the number of school user share has a weak correlation with personality traits as well (shown in Figure 3). There seems to be a positive correlation with Conscientiousness, Agreeableness, Neuroticism and Openness, and a negative correlation with Extroversion.

One of the most unusual correlations we find is between users who share five schools and those who share six. We notice that most of the users who share five schools are master students while those who share six are mainly master graduates who have full time jobs. The latter group declines sharply on Agreeableness and Openness, but they tend to become more conscientious.

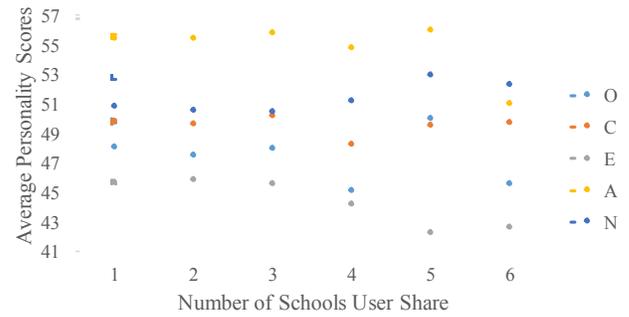

**Fig. 3.** The correlations between number of school user share and personality scores

### 5.4.3. Introduction information and personality

Users can optionally give a brief introduction (limited to 70 words) about themselves on Sina Weibo. There are 5,615 users have a self-introduction, counting for 86.83% in total. According to our study, users who introduce themselves score a little bit higher in Conscientiousness and Agreeableness.

Furthermore, Conscientiousness and Extroversion scores increase with the length of introduction, while Neuroticism scores decline evidently. Openness and Agreeableness scores remain stable (shown in Figure 4).

The result suggests that users who have detailed self-introductions tend to be more conscientious and extraverted, willing to be understood by more people.

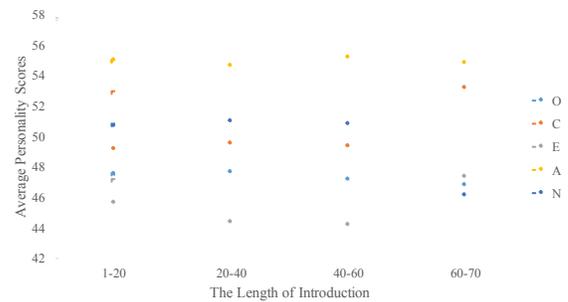

**Fig. 4.** The correlations between length of introduction and personality scores

### 5.4.4. User location and personality

There are 92.8% of users who share their location information. According to our study, users who share where they are usually score higher in Openness, Conscientiousness and Agreeableness.

To visualize the personality scores, we draw them on a map of China, with colors correspond to the level of value from various provinces. Figure 5 displays the average Neuroticism scores for different provinces. Geographically, users from southeastern coastal provinces usually score higher in Neuroticism.

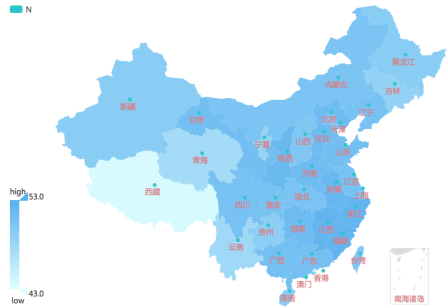

**Fig. 5.** The heatmap of personality scores from various provinces

### 5.5. Correlations between user's emoticon usage and personality

We study the emoticon usage difference between opposite personalities. We filter out emoticons that used more than 500 times, and calculate their respective proportions on each personality dimension. Some extremely popular emoticons, such as [心](❤️), [笑 cry](😂), [doge](🐶), [爱你](😘), [哈哈](😁) and [拜拜](🙂) display no great differences on antagonistic personality polarity in our study. Table 8 presents the typical emoticons which differ significantly in users' emoticon usage with opposite personality.

**Table 8.** Typical emoticons which differ significantly in users' emoticon usage with opposite personality

| Dimension | Frequently used emoticons |
|---|---|
| O- high | |
| O - low | |
| C- high | |
| C - low | |
| E - high | |
| E - low | |
| A- high | |
| A - low | |
| N- high | |
| N - low | |

## 6. CONCLUSIONS AND FUTURE WORK

Our work has proposed and demonstrated a promising psychological modeling approach of user portrait based on posted microblog texts and other social behaviors of users. Results of calculation analysis have shown that linguistic styles, self-description tags, emoticons usage and other demographic information can be used to compute their personality profiles without customer disruption.

As a follow-up work in the future, different online data across various social network platforms will be synthesized including Netease Music, Taobao and other media which share the same login accounts. Based on the trained psychological computational model, more clues on relationships between personality profiles of users and their music tastes and shopping habits can be discovered.

## 7. REFERENCES

[1] Cao, Belinda. "Sina's Weibo outlook buoys Internet stock gains: China overnight," *Bloomberg*, Dec. 2012.

[2] Zhao N, Jiao D, Bai S, et al. "Evaluating the Validity of Simplified Chinese Version of LIWC in Detecting Psychological Expressions in Short Texts on Social Network Services," *Plos One*, vol. 11, no. 6, 2016.

[3] Lin Qiu, et al. "Personality expression in Chinese language use," *International Journal of Psychology*, vol. 52, no. 3, pp. 463-472, 2017.

[4] Golbeck, Jennifer, C. Robles, and K. Turner. "Predicting personality with social media," *DBLP*, pp. 253-262, 2011.

[5] Sumner, Chris, et al. "Predicting Dark Triad Personality Traits from Twitter Usage and a Linguistic Analysis of Tweets," *International Conference on Machine Learning and Applications IEEE*, vol. 2, pp. 386-393, 2012.

[6] Hung, Chia Chuan, et al. "Tag-based user profiling for social media recommendation," *AAAI Workshop - Technical Report*, 2008.

[7] Pennebaker, James W., et al. "The development and psychometric properties of LIWC2015," 2015.

[8] Kalghatgi, Mayuri Pundlik, Manjula Ramannavar, and Nandini S. Sidnal. "A neural network approach to personality prediction based on the big-five model," *International Journal of Innovative Research in Advanced Engineering (IJIRAE)*, vol. 2, no. 8, pp. 56-63, 2015.

[9] Tausczik, Yla R., and James W. Pennebaker. "The psychological meaning of words: LIWC and computerized text analysis methods," *Journal of language and social psychology*, vol. 19, no. 1, pp. 24-54, 2010.

[10] G Gao R, Hao B, Li H, et al. "Developing simplified Chinese psychological linguistic analysis dictionary for microblog," *International Conference on Brain and Health Informatics*. Springer, Cham, pp. 359-368, Oct. 2013.

[11] Huang, Chin-Lan, et al. "The development of the Chinese linguistic inquiry and word count dictionary," *Chinese Journal of Psychology*, 2012.